\documentclass{article}

\usepackage{arxiv}

\usepackage{amsmath}
\usepackage[utf8]{inputenc} 
\usepackage[T1]{fontenc}    
\usepackage{hyperref}       
\usepackage{url}            
\usepackage{booktabs}       
\usepackage{amsfonts}       
\usepackage{nicefrac}       
\usepackage{microtype}      
\usepackage{lipsum}
\usepackage{fancyhdr}       
\usepackage{graphicx}       
\graphicspath{{media/}}     
\usepackage{doi}

\title{Harnessing Optical Imaging Limit through Atmospheric Scattering Media}

\author
{
	Libang Chen, Jun Yang, Lingye Chen, Yikun Liu \\
	School of Physics and Astronomy \\
	Sun Yat-Sen University \\
	Zhuhai 519082
	\texttt{\{Yikun Liu\}liuyk6@mail.sysu.edu.cn} \\
	\And
	Yuyang Shui, Jianying Zhou \\
	School of Physics \\
	Sun Yat-Sen University \\
	Guangzhou 510275
}
	
\renewcommand{\shorttitle}{\textit{arXiv}}

\begin{document}
\maketitle

\begin{abstract}
	Recording and identifying faint objects through atmospheric scattering media by an optical system are fundamentally interesting and technologically important. In this work, we introduce a comprehensive model that incorporates contributions from target characteristics, atmospheric effects, imaging system, digital processing, and visual perception to assess the ultimate perceptible limit of geometrical imaging, specifically the angular resolution at the boundary of visible distance. The model allows to reevaluate the effectiveness of conventional imaging recording, processing, and perception and to analyze the limiting factors that constrain image recognition capabilities in atmospheric media. The simulations were compared with the experimental results measured in a fog chamber and outdoor settings. The results reveal general good agreement between analysis and experimental, pointing out the way to harnessing the physical limit for optical imaging in scattering media. An immediate application of the study is the extension of the image range by an amount of 1.2 times with noise reduction via multi-frame averaging, hence greatly enhancing the capability of optical imaging in the atmosphere. 
\end{abstract}


\section{Introduction}

Imaging acquisition and recognition abilities are traditionally constrained by either the diffraction limit or the resolving capability of an optical imaging system. In low-visibility environments, optical images suffer severe contrast degradation due to atmospheric effects, resulting in reduced viewing distance and Angular Resolution (AR), hence the image perceptibility can be partially or completely impaired. Though existing imaging models, which are based on the contrast threshold of human vision \cite{ref1,ref2,ref3,ref4,ref5,ref6}, have made success in predicting the probability of human recognition of targets captured by traditional imaging systems through the atmosphere, with the evolution of imaging technologies including well-designed optical systems \cite{ref7,ref8,ref9}, high dynamic range sensors, sophisticated signal processing \cite{ref10,ref11,ref12,ref13}, and advanced displaying systems, the capability to perceiving images can be substantially enhanced. Hence there is a necessity to re-examine the optical imaging model in modern experimental settings.

In the context, this work introduces a reformulated imaging model taking into account all of the processes in imaging including optical transferring, recording, signal processing and perception. The model is based on the principles of the Meteorological Optical Range (MOR). Specifically, the role of a special parameter $k$, describing the image perceptibility, as introduced in the literature \cite{ref14}, is re-examined and its validity for discernible image is verified. By systematically exploring the imaging limits within the realm of geometric optical imaging systems, this work provides physical insights for dehazing algorithms \cite{ref15,ref16} and guidance for refining optical imaging systems. Our analysis was examined through experiments in a fog chamber as well as in outdoor settings. Good agreement between theoretical analysis and experimental results was obtained. Hence, the work descried here allows for the quantitatively determination of the physical boundary of the optical imaging in atmospheric scattering media.

\section{Principles and Methods}
\label{sec:P&M}

\subsection{The Imaging Model}
The Modulation Transfer Function (MTF) is widely used to describe the characteristics of an optical system \cite{ref17} and it is applied in this work to describe the imaging transferring behavior in atmospheric scattering media. Inspired by the theories of MOR, that the imaging perceptibility is determined by the vision's minimum modulation difference, we advocate for a refined definition of imaging model that relies merely on the appearance of the final modulation determined by the imaging system. Hence, the following formulation \cite{ref14} is applied to describe the imaging system:

\begin{equation}
    m_{d}MTF_{sys} = m_{d}^{img} \geq {\max\left\{ {km_{d}^{noise},\frac{1}{\Gamma}} \right\}},
    \label{eq1}
\end{equation}

Where $m_{d}$, $m_{d}^{img}$, and $m_{d}^{noise}$ represents the modulation (or contrast) of the target, the imaging system's output, and the noise respectively.  $MTF_{sys}$ is the system's MTF, $k$ is the perceiving factor, a multiplier of the theoretical minimum discernible modulation, and $\Gamma$ is the minimum of the sensor's dynamic range and bit-depth. Eq. \eqref{eq1} presents two critical conditions: the Signal-to-Noise Ratio (SNR) condition, delineating the system's capability to distinguish modulation variations across spatial frequencies amid noise; and the Signal-to-Interference Ratio (SIR) condition, illustrating an absolute threshold for modulation detection limited by the system's sensitivity to optical intensity. 

Specially, for the SNR condition, the parameter $k$ serves as a comprehensive indicator of the effectiveness of image perceptibility, spanning from naked eye observation to algorithm-assisted analysis, with its magnitude reflecting the efficacy of post-processing techniques. When the imaging modulation falls below $km_{d}^{noise}$, representing the system perceptibility, the target becomes obscured by noise, thus reaching the noise-dominated imaging limit. Likewise, when signal modulation falls short of $\Gamma$, the signal is lost during the optoelectronic recording and conversion.

The expression for $m_{d}$ is as follows:

\begin{equation}
    m_{d} = \frac{I_{W} - I_{B}}{I_{W} + I_{B}} = \frac{r_{W} - r_{B}}{r_{W} + r_{B}},
    \label{eq2}
\end{equation}

Where $I_{W}$ and $I_{B}$ represent the maximum and minimum light intensities, while $r_{W}$ and $r_{B}$ represent the maximum and minimum reflectance.

The light transmission and imaging process involves the ambient light $A$, the reflected light $I_{r}$, the interference light $I_{it}$ and the scattered light $I_{sa}$. The latter three transmit through the atmosphere and lenses and are then converted into electrical signals by the imaging sensor. Thus, the MTF of the imaging system, $MTF_{sys}$, can be expressed as follows:

\begin{equation}
    MTF_{sys} = ~MTF_{A}MTF_{L}MTF_{S},
    \label{eq3}
\end{equation}

Where $MTF_{A}$, $MTF_{L}$, and $MTF_{S}$ denote the MTFs of the atmosphere, lens, and sensor respectively. The $MTF_{A}$ is divided into three components \cite{ref18}: atmospheric scattering $MTF_{SA}$, atmospheric absorption $MTF_{AS}$, and atmospheric turbulence $MTF_{TU}$. It was proven in \cite{ref19,ref20,ref21} that $MTF_{SA}$ is approximately a constant at high frequencies. In a thermally stable environment, the turbulence MTF can be neglected \cite{ref22}. Consequently, the major effect originates from $MTF_{AS}$. With Appendix, the expressions for all the MTFs can be written as:

\begin{align}
    MTF_{A} &= \frac{1}{2{\exp(\tau)} - 1},\\
    MTF_{L} &= {\exp\left\lbrack {- \frac{\pi^{2}\delta^{2}}{4{\ln(2)}}\nu^{2}} \right\rbrack},\\
    MTF_{s} &= {~exp}\left\lbrack {- 2\pi^{2}\left( \frac{L_{s}}{6} \right)^{2}\nu^{2}} \right\rbrack,
    \label{eq4-6}
\end{align}

Where $\tau$ represents the optical thickness, $\delta$ is the Airy disk diameter, $L_{s}$ is the line spread width of the sensor, and $\nu$ is the spatial frequency. 

The modulation of noise \cite{ref14} can be calculated as follows:

\begin{equation}
    m_{d}^{noise} = ~\frac{< \left| {\mathcal{F}(n)} \right| >}{< n > ~},
    \label{eq7}
\end{equation}

Where $\mathcal{F}( \cdot )$ signifies the Fourier transform, $< \cdot >$ denotes taking the mean value, and $n$ represents the noise from the system, originating both from the atmosphere and the imaging system itself.

By substituting the above equations to Eq. \eqref{eq1}, we can achieve the expression for AR as a function of $\tau$ in the SNR condition:

\begin{align}
    &\alpha = \frac{\pi^{4}\left( {\frac{\delta^{2}}{ln2} + \frac{2}{9}L_{s}^{2}} \right)}{\ln\left\{ {\frac{k}{m_{d}}\frac{< \left| {\mathcal{F}(n)} \right| >}{< n > ~}\left\lbrack {2{\exp(\tau)} - 1} \right\rbrack} \right\}},\label{eq8}\\
    &AR = 2{\arctan\left( \frac{\sqrt{\alpha}}{f} \right)},\label{eq9}
    \end{align}

Where $\alpha$ is the minimum distinguishable size. To satisfy the SIR condition, there exists a maximum value for $\tau$:

\begin{equation}
    \tau_{max} = ~ - ln2 + {\ln\left( {\frac{m_{d}}{\Gamma} + 1} \right)},
    \label{eq10}
\end{equation}

Eq. \eqref{eq8} indicates that, when attenuation is weak, the limitation of the imaging is entirely due to Rayleigh criterion and imaging resolution. The above two equations quantify two boundaries for imaging, representing the imaging limit, revealing the complex nature of imaging through atmospheric scattering media. The discussion indicates that the imaging limit can be harnessed via various strategies, such as refining imaging methods, employing lens with higher numerical apertures, enhancing sensor dynamic range and bit-depth, reducing system noise, and implementing advanced algorithms.

The choice of $k$ value is pivotal in framing our approach to quantifying imaging limitations through atmospheric scattering media. As indicated by \cite{ref23}, within the human eye sensitivity (10-bit) \cite{ref24,ref25}, there exists a positive correlation between the probability of visual perception and the value of $k$ for unprocessed images. When $k=5.5$, the human eye can reliably detect objects. Following the guidelines of the International Union of Pure and Applied Chemistry (IUPAC), we set $k=3$ as the detection limit \cite{ref26,ref27}. Drawing from both the theoretical groundwork laid out in prior discussion and the empirical benchmarks set by existing research, the experimental validation of $k$ threshold becomes imperative, as it offers a critical test of the applicability in real-world application.

\subsection{Experimental setup}

The objective of the indoor experiment was to rigorously evaluate model's accuracy in predicting imaging limits within a controlled low-visibility environment, and to replicate typical imaging scenarios examining both the SIR condition and the SNR condition (whether $k=3$). Thus, we chose the following two cameras: Daheng MER-232-48GM-P NIR (8-bit) and PCO EDGE 4.2 (16-bit). MER-232-48GM-P NIR is optimized for near-infrared and is SIR limited, whereas the EDGE 4.2 is visible light responsive and SNR limited. Lens choices were the Kowa M7528-MP and Zeiss Milvus 2/100M with the focal lengths of 75 mm and 100 mm respectively to match the camera's pixel physical dimensions. 

The fog chamber experiment was orchestrated at the Visibility Calibration Laboratory of the China Meteorological Administration in Shanghai, China. The target employed was the USAF 1951 resolution test chart, featuring 6 groups (G:0 to G:5) and 6 elements (E:1 to E:6) within each group, with the test chart's reflective area for the black portion around 2.5\% and for the white area around 90\%. Alignment between the target and cameras is horizontal, with a working distance of 16 meters. A visibility meter (HUAYUN SOUNDING DNQ1) with 10\% of error was placed at the same height as the target to obtain the visibility \cite{ref28,ref29} as Fig. ref{fig1} demonstrates.

\begin{figure}[htbp]
    \centering
    \includegraphics[width=\columnwidth]{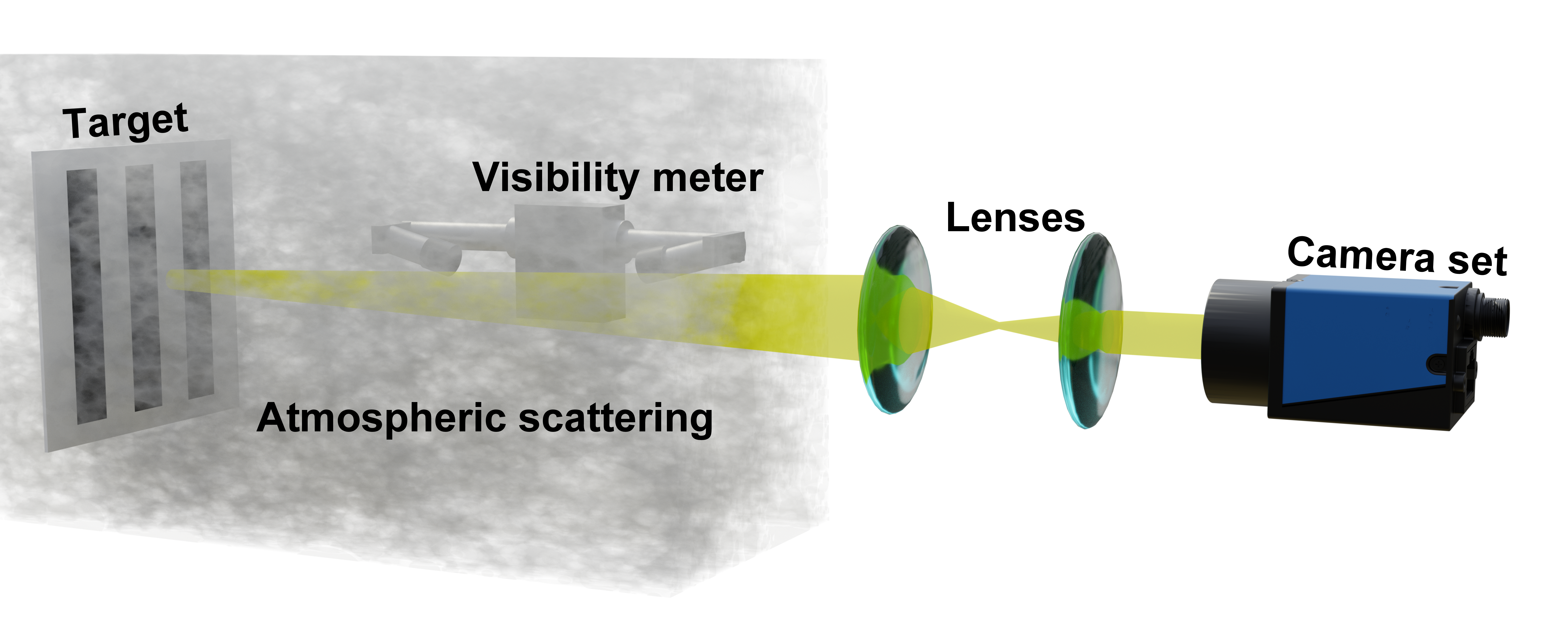}
    \caption{Schematic of the optical imaging process through atmospheric scattering media under external illumination.}
    \label{fig1}
\end{figure}

In pursuit of further advancements in the system's ability to penetrate fog and to validate the SNR condition, we conducted an in-depth test and analysis in outdoor settings. According to Eq. \eqref{eq8}, noise reduction can effectively improve the system's imaging capability. We conducted the simple approach of Multi-Frame (M-F) averaging to reduce Gaussian noise \cite{ref30, ref31}, thereby increasing the SNR. The noise variance can present an inverse proportional decay with the number of frames averaged. The noise variance is defined by:

\begin{equation}
    Varience = < \left( {n - < n >} \right)^{2} >,
    \label{eq13}
\end{equation}

Where we identified and selected a homogeneous region within the image and filter it with a moderate kernel to obtain $n$. In the outdoor experiment, we utilized the PCO EDGE 4.2, paired with a Canon EF 400 mm F/5.6L lens. The resolution test chart was positioned 5 kilometers away from the camera set. Visibility measurements were facilitated by a drone-mounted visibility meter, offering real-time data on atmospheric conditions. The data were sampled at intervals of 100 milliseconds and 200 images were averaged. 

\subsection{Experimental Imaging Criteria}

The recorded images were further processed with the following technique. For the 8-bit images, the series was shown by: the original image, Dark Channel Prior (DCP) \cite{ref30}, and Clipped Limit Adaptive Histogram Equalization (CLAHE) \cite{ref31}. The DCP leverages atmospheric degradation model to enhance images. The CLAHE enhances local modulation by dividing the image into small regions and applying histogram equalization separately to each region. Both are effective dehazing algorithms for enhancing images. For the 16-bit data, participants were invited to perform custom histogram adjustments to achieve the optimal enhancement algorithm without prior knowledge of the target.

\begin{figure}[htbp]
    \centering
    \includegraphics[width=\columnwidth]{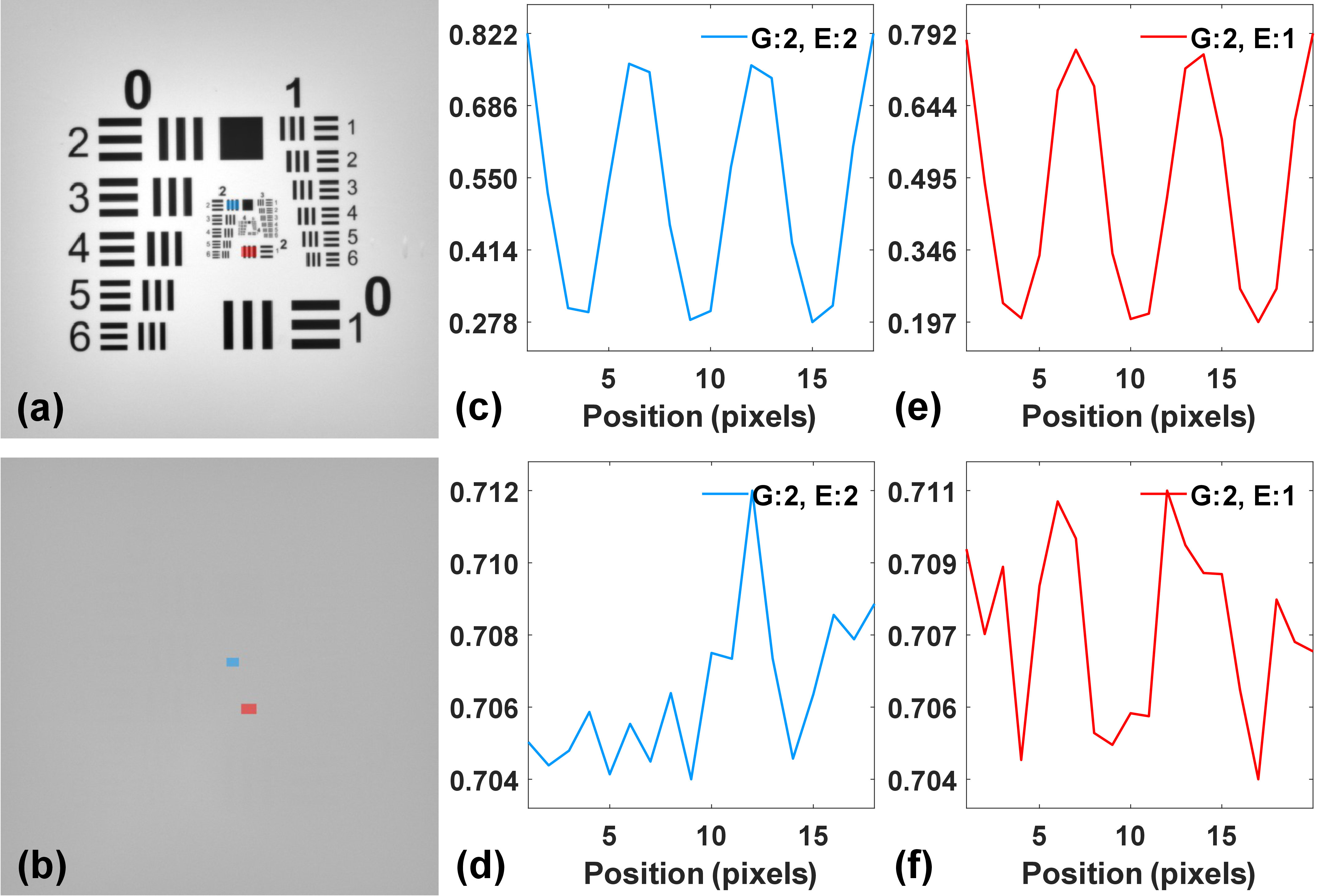}
    \caption{Original Data from fog chamber experiment. Targets in (a) high and (b) low visibility conditions, along with the average normalized values for the red region in (c) high and (d) low visibility, and the blue region in (e) high and (f) low visibility.}
    \label{fig2}
\end{figure}

Examples of the data collected in the fog chamber experiment are displayed in Fig. \ref{fig2}. The figure demonstrates the target alongside the average and normalized values inside the red and blue regions in Fig. \ref{fig2}(a) and (b) with respectively high and low visibility. In Fig. \ref{fig2}(a), the imaging is limited by the system's resolution, extending beyond G:2, E:2. Conversely, in Fig. \ref{fig2}(b), the imaging is limited by visibility, thus three clear peaks are discernible before G:2, E:2, and no discernible feature beyond. The quantitative analysis is in good agreement with the human visual observation as well as algorithm-aided observation, indicating that the obliteration of signal features renders the task of discernment unfeasible without prior knowledge of the target. These findings not only delineate the limits for image perceptibility but also set the stage for study to harness the resolution limits.

\section{Results and Discussion}

\subsection{Fog Chamber Experiment}

\begin{figure}[htbp]
    \centering
    \includegraphics[width=\columnwidth]{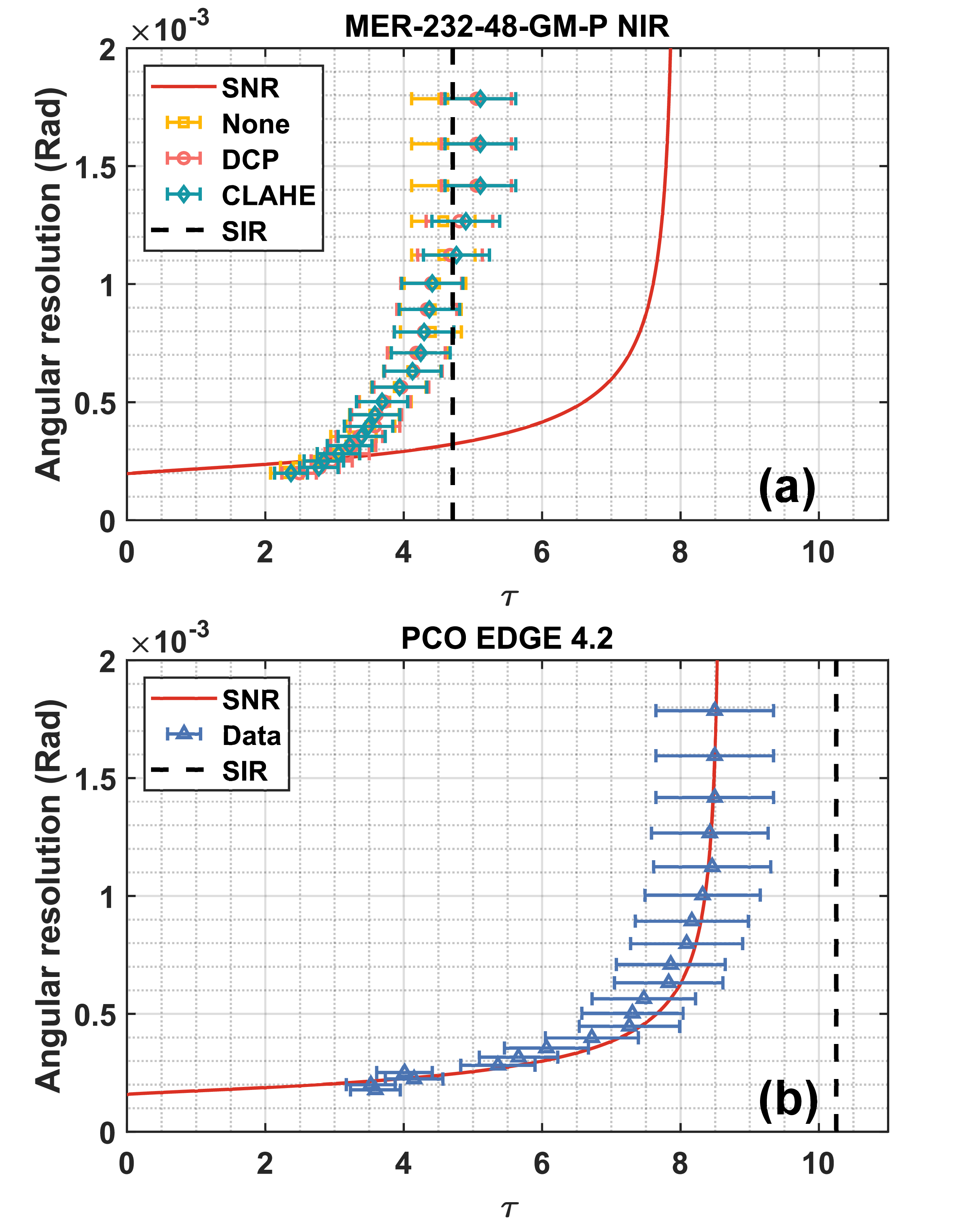}
    \caption{Relationship between AR and $\tau$, Prediction versus Data, (a) MER-232-48-GM-P NIR and (b) PCO EDGE 4.2.}
    \label{fig3}
\end{figure}

For the MER-232-48GMP NIR imaging system, Fig. \ref{fig3}(a) presents the data analysis where the red curve represents the SNR condition, and the black dashed line signifies the SIR condition. The yellow, orange, and blue error points respectively correspond to the data observed without algorithms, using DCP, and using CLAHE for the corresponding minimum AR. With algorithm-aided vision, the $\tau$ penetrable by the imaging system improves approximately 12\%. However, in terms of smaller objects, these algorithms offer a relatively minor enhancement for $\tau$. Here, the imaging limit is mainly determined by the SIR condition.

For the PCO EDGE 4.2 imaging system, Fig. \ref{fig3}(b) illustrates the performance analysis where the red curve and black dashed line respectively represent the SNR and the SIR conditions. The blue error points represent the observed data. Given the system is SNR limited, we focus on fitting the data with respect to $k$. To quantify the fitting's accuracy and reliability, we employed the statistical measures: the R-squared ($R^2$) and root mean square error (RMSE). Their expressions are as follows:

\begin{align}
    &R^{2} = 1 - \frac{\Sigma_{i = 1}^{data}\left( {y_{i} - y_{i}^{pred}} \right)^{2}}{\Sigma_{i = 1}^{data}\left( {y_{i} - < y >} \right)^{2}},\\
    &RMSE = \sqrt{\frac{1}{data}\Sigma_{i = 1}^{data}\left( {y_{i} - y_{i}^{pred}} \right)^{2}},
    \end{align}
\label{eq11-12}

Where $y$ is the measured value, $y^{pred}$ is the predicted value, and $data$ is the number of data. The fitted $k$ is approximately 3, aligning well with the IUPAC limit, with an R-squared of 0.9544, and RMSE 0.0001, affirming the imaging limit's reliance on SNR conditions. 

These results underscore the versatility and accuracy of our proposed model in predicting the Single-Frame (S-F) imaging limit of the detecting system affected by atmospheric scattering. The experiment not only confirms the model's applicability across different camera systems and conditions but also highlights the potential for enhancement algorithms to mitigate the visibility limitations in foggy environments.

\subsection{Outdoor Experiment}

We first examine the relationship between the modulation and variance of the image noise. $m_{d}^{noise}$ is measured to examine the relationship between the image clarity with the number of averaging and the variance is measured to confirm the elimination of Gaussian noise. The numerical results of the two quantities calculated based on measured experimental data are illustrated in Fig. 4(a) and (b), showcasing the decay of variance and $m_{d}^{noise}$ in accordance with the law associated with Gaussian noise. When the number of averaging is larger than 100, $m_{d}^{noise}$ and variance quickly converge to constant level, indicating that the image contains Gaussian noise and non-Gaussian noise that cannot be eliminated by averaging \cite{ref34}. 

\begin{figure}[htbp]
    \centering
    \includegraphics[width=\columnwidth]{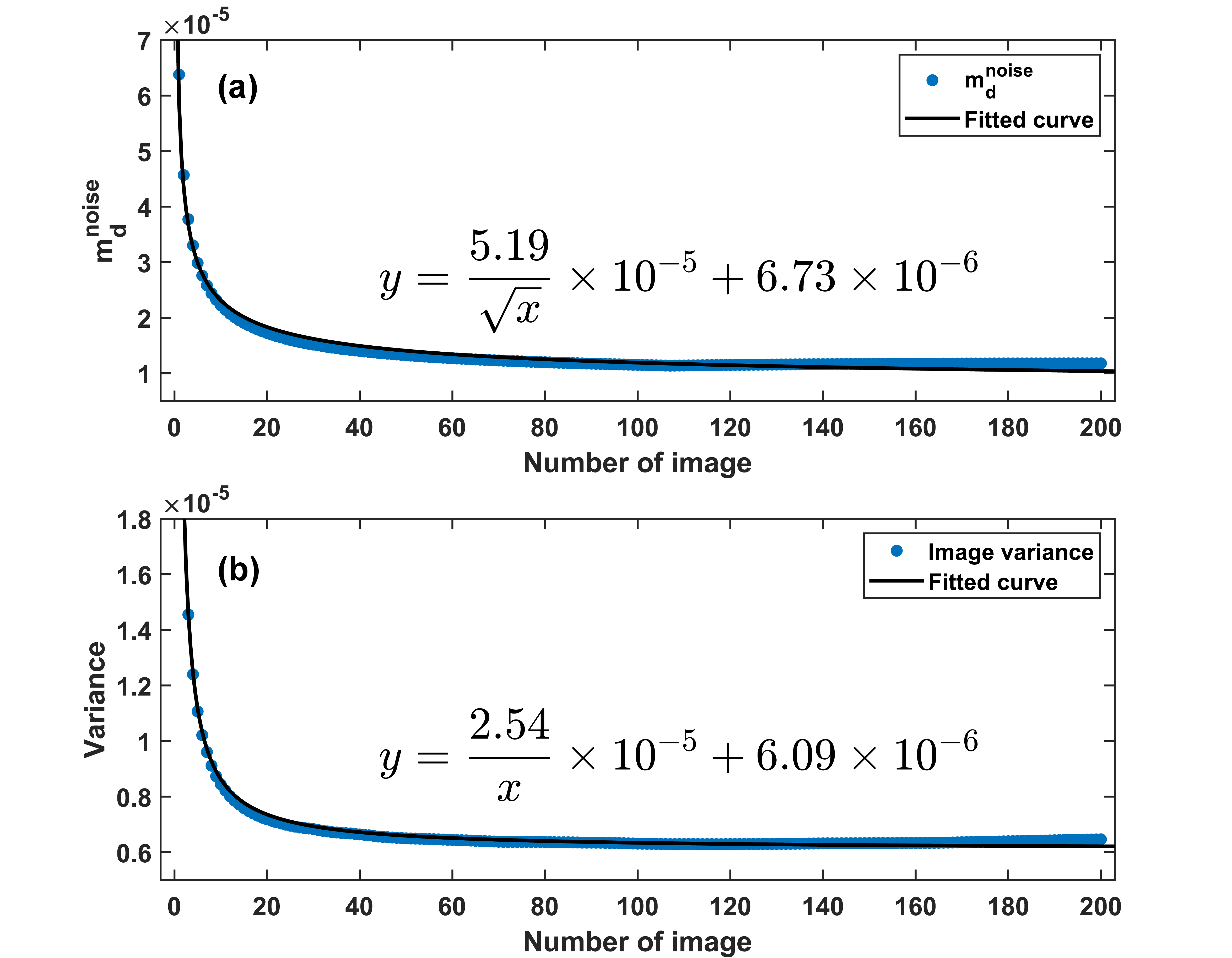}
    \caption{Relationship between number of averaging and (a) image variance and (b) SNR.}
    \label{fig4}
\end{figure}

\begin{figure}[htbp]
    \centering
    \includegraphics[width=\columnwidth]{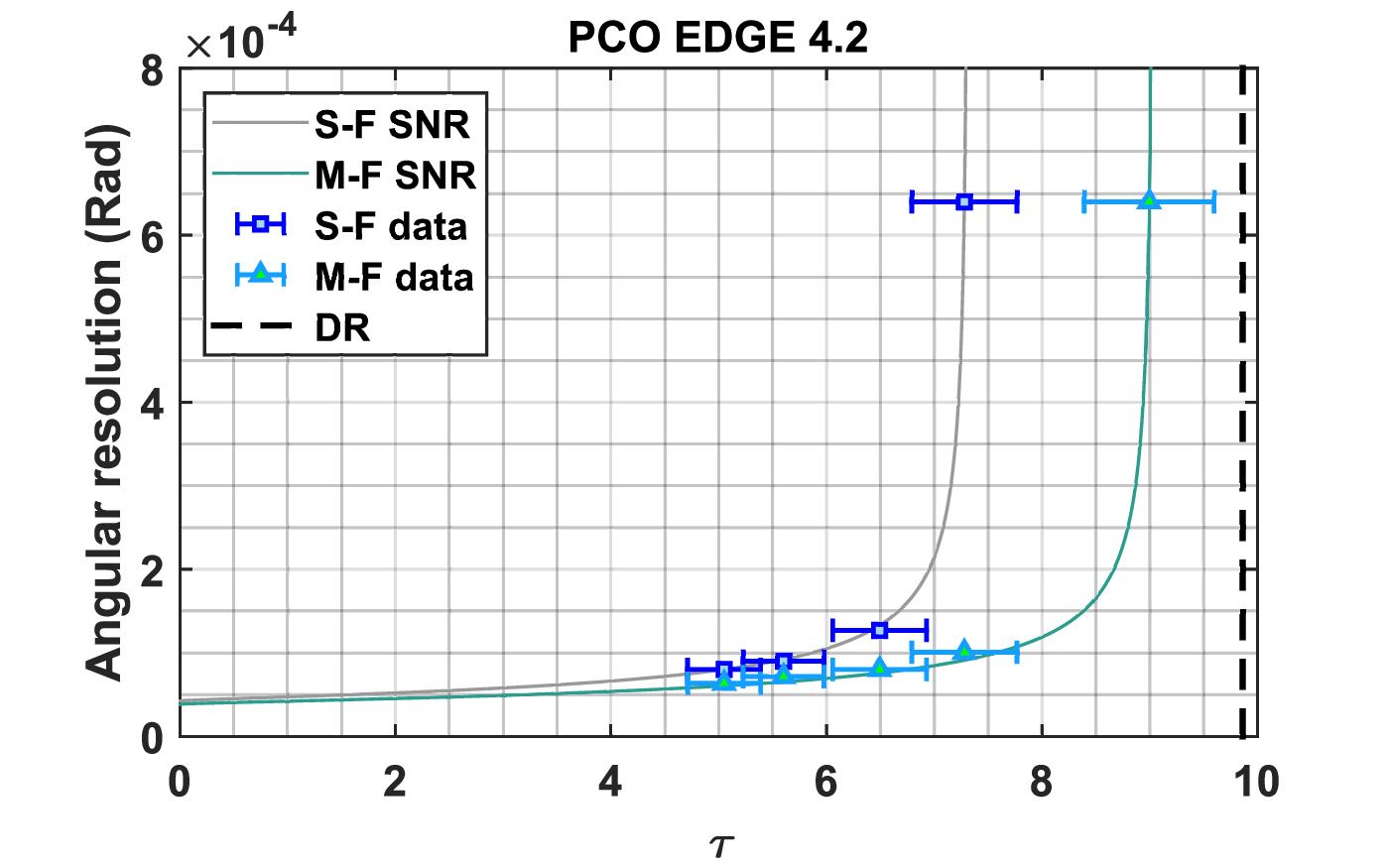}
    \caption{Relationship between AR and $\tau$ in field experiment.}
    \label{fig5}
\end{figure}

We then examine the efficacy of using averaging in actual imaging. Fig. \ref{fig5} depicts the AR for S-F and M-F averaged images against $\tau$, highlighting the impact on SNR. The R-square and RMSE values are 0.92761 and $7.2996\times10^{-5}$ for S-F fitting, and 0.94839 and $5.7092\times10^{-5}$ for M-F fitting, respectively. In the case of a camera limited by SNR, employing noise reduction through averaging has effectively extended the imaging range to 1.2 times that of the S-F imaging system, corresponding to an increase of 1.7 $\tau$. Comparing to the S-F system, where the $k$ value associated with its $m_{d}^{noise}$ is 3, the M-F system exhibits a significantly reduced $k$ value at 0.1839, underscoring the technique's potential to extend the imaging system's operational $\tau$. 

As a result, the field experiment reinforces the comprehensive framework of our proposed model, highlighting both the potential and limitations of current image enhancement techniques in addressing atmospheric scattering effects.

\section{Conclusion}
In conclusion, we derived a comprehensive physical model to describe the behavior of optical images after traversing a specific optical thickness in the atmosphere based on the principles of MOR. We show explicitly that the image can be retrieved with the requirement that the image modulation survive to a minimum level so that the system detection allows the modulation to be detected via high dynamic range detectors or alternatively via multi-frame averaging. Experimental validations of our prediction were conducted in both a fog chamber and outdoor settings, revealing a substantial alignment, demonstrating that the optical image limit can be harnessed to achieve its best performance via optimized optical system, suitable detecting system, and effective post-signal processing.

\section{Acknowledgments}
This work was supported by the National Natural Science Foundation of China (61991452, 12074444); the National Key Research and Development Program of China (2022YFA1404300, 2020YFC2007102); and the Guangdong Major Project of Basic and Applied Basic Research (2020B0301030009). The authors acknowledge Yuyang Shui for helpful discussion.

\section{Appendix}

\title{Appendix: The System MTF}

In this section, we focus on the derivations of the system MTF.

\begin{figure}[htbp]
    \centering
    \includegraphics[width=\columnwidth]{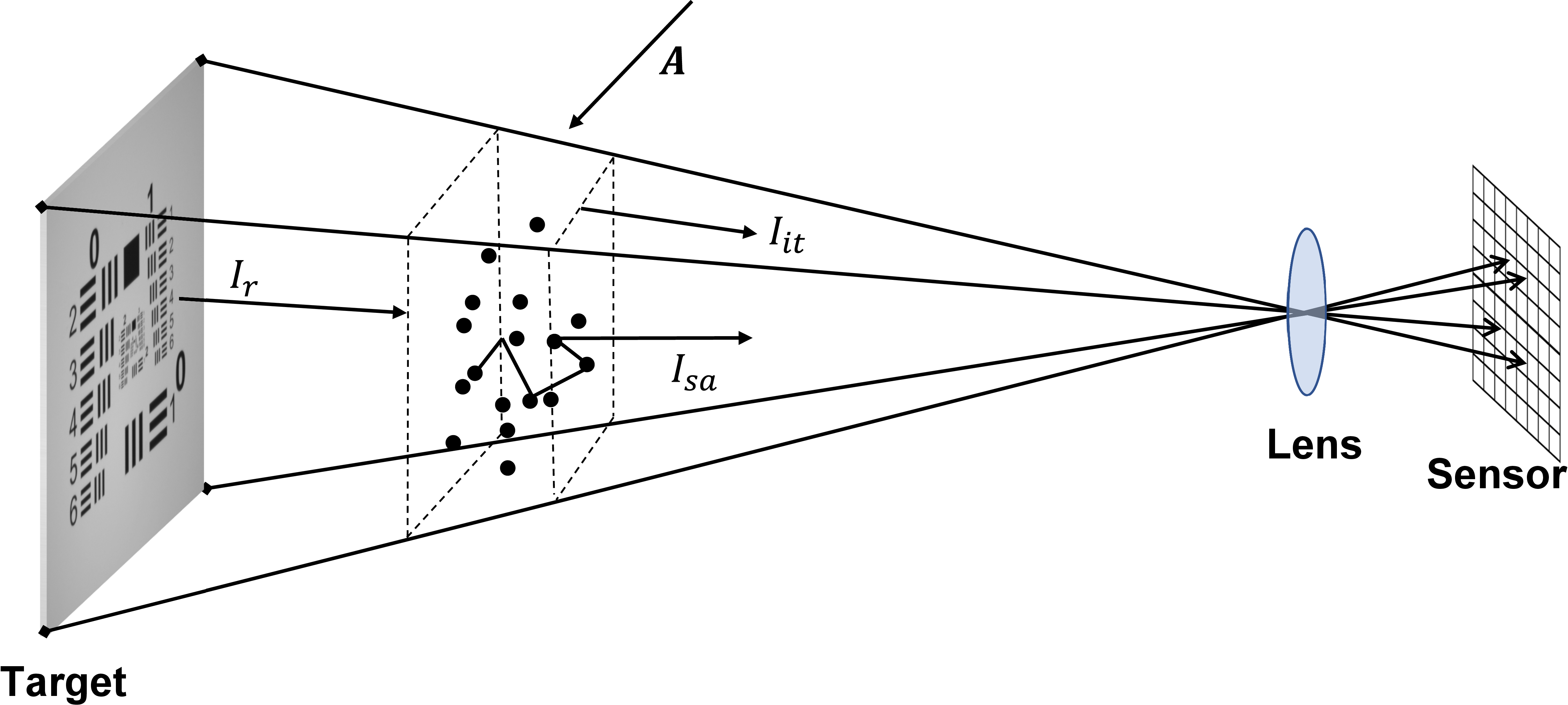}
    \caption{Schematic of the optical imaging process through atmospheric scattering media under external illumination.}
    \label{fig6}
\end{figure}

The optical imaging process is demonstrated by Fig.\ref{fig6}. The ambient light $A$ illuminating the target undergoes diffuse reflection, resulting in the reflectance within the field of view, denoted as  $r(x,y)$. The imaging light consists of the reflected light $I_{r}= Ar(x,y)$ as well as the interference light $I_{it}$ and scattered light $I_{sa}$. When the target is of restricted AR, the $I_{r}$ enters the lenses at a fixed angle. Therefore, we consider the extinction coefficient $\beta_{ext}$ to be a constant. This allows us to formulate the radiative equation as follows \cite{ref35}:

\begin{equation}
    \begin{split}
        \frac{dI_{r}\left( {x,y,z} \right)}{dz} &= -\beta_{ext}I_{r}\left( {x,y,z} \right) \\
        &\quad + \beta_{sca}A + \Delta I_{sa}(x,y,dz),
    \end{split}
\label{eq14}
\end{equation}

Where $\beta_{sca}$ is the scattering coefficient. The coordinates $x$, $y$, $z$ are established in a reference system with the target as the origin. Assuming the distance between the target and the lenses is $L$, we denote $\tau$ as $\beta_{ext} L$.

The light intensity distribution can be calculated as follows:
\begin{equation}
    \begin{matrix}
        {I_{r}\left( {x,y,L} \right) = {\exp\left( {- \tau} \right)}\left\lbrack {I_{r}\left( {x,y,0} \right) - \frac{\beta_{sca}A}{\beta_{ext}}} \right\rbrack} \\
        {+ \beta_{sca}A/\beta_{ext}~+ I_{sa}(x,y,L),}
    \end{matrix}
    \label{eq15}
\end{equation}

Substituting Eq \eqref{eq15} into the definition of modulation, we obtain the relationship between modulation depth and $\tau$:

\begin{equation}
    m_{d}(\tau) = \frac{m_{d}}{1 + \frac{2\beta_{sca}}{\beta_{ext}\left( {r_{W} + r_{B}} \right)}\left\lbrack {\exp(\tau)} - 1 \right\rbrack},
    \label{eq16}
\end{equation}

For ideal target imaging with an infinitely distant background and an object appearing as absolute black, $r_{B}=0$ and $r_{W}=\beta_{sca}/\beta_{ext}$, This leads to a simplified expression for modulation depth induced by atmospheric attenuation with respect to $\tau$:

The Point Spread Function (PSF) of the lenses is characterized by Bessel functions, and its diameter, where the first minimum occurs, is known as the Airy disc. The full width at half-maximum (FWHM) of the Airy disc is given by:

\begin{equation}
    \delta = \frac{1.025\lambda}{D}f,
    \label{eq17}
\end{equation}

Where $\delta$ is the Airy disc diameter, and $D$ is the effective diameter of the lenses. The lenses' PSF can be approximated as a Gaussian function \cite{ref36}

\begin{equation}
    PSF_{L}(r) = {\exp\left[{- \frac{r^{2}}{2\left( \delta/2\sqrt{2ln2} \right)}} \right]},
    \label{eq18}
\end{equation}

Thus, the MTF of the lens can be expressed as:

\begin{equation}
    MTF_{L} = {\exp\left [{- \frac{\pi^{2}\delta^{2}}{4{\ln(2)}}\nu^{2}} \right]},
    \label{eq19}
\end{equation}

Where $\nu$ is the spatial frequency. In theory, the MTF of the sensor can be simulated by a sinc function \cite{ref37}, defined as ${\sin\left( {\pi\nu} \right)}/(\pi\nu)$, but in practical engineering, a Gaussian function also provides a good approximation \cite{ref38,ref39}. The PSF of the sensor can be obtained by applying low-pass filtering to the Edge Spread Function (ESF) derived from the sensor's captured edges, followed by a Fourier transform of the Line Spread Function (LSF). Therefore, its expression is:

\begin{equation}\
    PSF_{S} = {\exp\left[{- \frac{r^{2}}{2\left( {L_{s}/6} \right)^{2}}} \right]},
    \label{eq20}
\end{equation}

Where $L_{s}$ is the total width of the LSF. Consequently, the MTF of the sensor can be written as:

\begin{equation}
    MTF_{S} = {\exp\left[{- 2\pi^{2}\left( {L_{s}/6} \right)^{2}\nu^{2}} \right]},
    \label{eq21}
\end{equation}

\bibliographystyle{unsrt}  
\bibliography{manuscript}

\end{document}